\documentclass[lineno]{jfm}

\usepackage{graphicx}
\usepackage{newtxtext}
\usepackage{newtxmath}
\usepackage{natbib}
\usepackage{hyperref}
\usepackage{caption}
\hypersetup{
    colorlinks = true,
    urlcolor   = blue,
    citecolor  = black,
}

\newcommand{\RomanNumeralCaps}[1]

\newcommand{\bU}{{\boldsymbol{{\cal U}}}}
\newcommand{\bP}{{\boldsymbol{{\cal P}}}}
\def\la{\left \langle}
\def\ra{\right \rangle}
\providecommand\bnabla{\boldsymbol{\nabla}}

\def\bOmega{ \boldsymbol{\Omega}}
\def\bxi{ \boldsymbol{\xi}}

\shorttitle{Coupled model for gravity wave-mean flow interaction}
\shortauthor{B. Gallet}

\title{Surface gravity wave–mean flow interaction with comparable spatial scales. Part II: two-way coupling and wave-wave interactions.}

\author{Basile Gallet  \corresp{\email{basile.gallet@cea.fr}}
}

\affiliation{\aff{1}{Universit\'e Paris-Saclay, CNRS, CEA, Service de Physique de l'Etat Condens\'e, 91191 Gif-sur-Yvette, France.}}

\begin{document}
\maketitle


\begin{abstract}
We consider narrow-band deep-water surface gravity waves propagating above a background flow whose spatial scale is comparable to the wavelength. We focus on the regime where the background-flow speed is comparable to the Stokes drift of the waves to derive a two-way coupled model between the fast waves and the slow background flow. Nonlinear wave interactions arise at the same asymptotic order as the two-way coupling and must therefore be included in the derivation. The resulting model consists of a nonlinear version of the reduced wave equation obtained in part I, coupled to a Craik-Leibovich equation governing the evolution of the background flow. The model describes the evolution of the wave field and background flow over the slow turnover frequency of the background flow, thus saving the computational burden of temporally resolving the fast wave frequency. The inviscid model exactly conserves wave action, mechanical energy and horizontal momentum, while the viscous model exactly conserves horizontal momentum. The coupling terms are cast in compact form for ease of physical interpretation and numerical implementation. 
\end{abstract}


\section{Introduction}


The interaction between surface waves and background flows goes both ways, with currents refracting or scattering the wave field, and waves generating or distorting background flows. In the Ocean, mesoscale vorticity refracts the swell propagating across ocean basins~\citep{kenyon1971wave,gallet2014refraction,ardhuin2017small}, while surface waves energize near-inertial waves~\citep{hasselmann1970wave,wagner2021near}, modify the Ekman spirals~\citep{gnanadesikan1995structure,polton2005role,seshasayanan2019surface} and conspire with vertically sheared background flows to induce Langmuir cells~\citep{craik1976rational,craik1977generation,leibovich1977convective,mcwilliams1997langmuir}, thereby contributing to upper-ocean mixing. In the laboratory, the two-way coupling between background flows and surface waves has also been investigated, with studies characterizing the impact of a background flow on the wave field~\citep{vivanco2000surface,vivanco2004experimental,cobelli2009global,gutierrez2016surface,humbert2017surface}, the distortion of a background flow by a wave field~\citep{teixeira2002distortion}, or both simultaneously~\citep{humbert2017wave}.
 
While the most common approach to study these effects is based on a scale separation assumption between the wavelength and the scale of the background flow~\citep{dysthe2001refraction,ardhuin2012numerical,Buhler2014,pizzo2021particle,Boury2023,tlili2026equilibrium,vanneste2026consistent},  such scale separation rarely arises in laboratory experiments or for Langmuir cells in the ocean. 
A similar absence of scale separation for the interaction between near-inertial waves and currents in the Ocean has spawned important theoretical developments. \citet{YBJ1997} derived a reduced equation governing the evolution of near-inertial waves modulated by a background flow with a comparable spatial scale~(see the subsequent studies by~\citet{balmforth1998enhanced,danioux2015concentration,asselin2020penetration,conn2024interpreting,Tlili2025}), while~\citet{Xie2015} and~\citet{wagner2016three} later extended this approach to include the feedback of the near-inertial wave field onto the background flow~(see also~\citet{kafiabad2021interaction}). Similar developments in the context of surface gravity waves are desirable, as the equilibrated state of oceanic Langmuir cells cannot be theoretically characterized  without a model that remains valid in the absence of scale separation~\citep{suzuki2019physical}. 

With this motivation in mind, in a companion paper (\citet{galletpart1}, part I in the following) we derived reduced equations governing the dynamics of surface gravity waves above a weak background flow, in the regime where the wavelength is comparable to the characteristic scale of the background flow. In this article, we complete the program by coupling the reduced wave equation to the Craik-Leibovich (CL) equation governing the evolution of the mean flow~\citep{craik1976rational}. 

In part I, the small parameter allowing for the reduction of the equations is the Froude number $\epsilon$ of the background flow. That is, the background flow is assumed to be much slower than the group velocity of the waves. The amplitude $\delta$ of the waves was assumed to be smaller than any power of $\epsilon$, and in that sense the waves were considered `infinitesimal'. By contrast, in the present article we consider waves of finite amplitude scaling as $\delta=\sqrt{\epsilon}$ and we derive a set of coupled, reduced equations governing the co-evolution of the waves and the background flow. As compared to the interaction between near-inertial waves and mean flows~\citep{Xie2015}, an additional complexity for the surface-wave system is that two-way coupling and nonlinear wave interaction arise at the same asymptotic order in the expansion. Asymptotic consistency dictates that such nonlinear four-wave processes be included in the model.

As we were finalizing the present two-part article, we became aware of the very recent preprint by \citet{onuki2026reduced}. Using the same scalings, they derived a two-way coupled model between surface gravity waves and background currents using the method of reconstitution~\citep{roberts1985introduction,wagner2017asymptotic,thomas2017new,thomas2018amplitude}, including finite depth and global rotation. However, they omit wave-wave interactions, resorting instead to a `phenomenological closure that neglects quartic wave-wave interactions'. 
The reduced wave equation we derive in part I of the present study has the advantage of being particularly compact.
In the present article, we further augment the reduced wave equation with the appropriate nonlinear terms, obtained through an asymptotic derivation, as such wave-wave interaction terms are required for asymptotic consistency of the coupled model. Additionally, we cast the feedback of the waves onto the background flow in terms of the solenoidal expression for the Stokes drift proposed by~\citet{vanneste2022stokes}. The result is a fully asymptotically consistent two-way coupled model, with coupling terms cast under a compact form, for ease of physical interpretation and numerical implementation.

The two-way-coupled model is introduced in section~\ref{sec:coupledmodel}. The model exactly conserves wave action, mechanical energy and horizontal momentum, as discussed in section~\ref{sec:conservation}. The asymptotic derivation of the model itself is deferred to sections~\ref{sec:derivation} and~\ref{sec:derivationNL}. Section~\ref{sec:derivation} focuses on the asymptotic derivation of two-way coupling. It consists of the asymptotic expansion leading to the reduced wave equation in part I, intertwined with the standard asymptotic expansion leading to the CL equation. Section~\ref{sec:derivationNL} focuses on the asymptotic derivation of the nonlinear wave-wave interaction terms entering the reduced wave equation. In section~\ref{sec:viscous} we include viscosity in the coupled model, a necessary step to implement the model in most numerical solvers. In the discussion section~\ref{sec:discussion} we briefly discuss a more `minimal' version of the model, where some terms are eliminated at the expense of making energy and momentum adiabatic invariants only, as opposed to exact invariants. We conclude in section~\ref{sec:conclusion}.


\section{The coupled model\label{sec:coupledmodel}}


We consider the flow of an incompressible fluid in a semi-infinite domain bounded from above by a free surface. The atmosphere above the fluid has low density and uniform pressure. The fluid motion consists of a three-dimensional background flow ${\bf U}({\bf x},t)$ together with small-amplitude deep-water surface gravity waves. The background flow is slow as compared to the group velocity of the waves, with a small Froude number $\epsilon$ and a negligible surface deformation. 
We consider narrow-band surface gravity waves propagating above the background flow, with a central angular frequency $\omega_0$. The equations are non-dimensionalized using $\omega_0$ and gravity $g$. The leading-order wave field thus consists of signal close to near-unit dimensionless angular frequency only, corresponding to near-unit dimensionless wavenumber.

In part I, we focused on infinitesimal waves, scaling the wave amplitude with a small parameter $\delta$ and assuming that $\delta$ was much smaller than any power of $\epsilon$. In the present study, we instead consider the regime $\delta=\sqrt{\epsilon}$. This scaling ensures that the Stokes drift of the wave field, arising at order $\delta^2=\epsilon$, is comparable to the speed of the background flow, in line with the Craik-Leibovich ordering.

The wave field is described in terms of a demodulated complex amplitude ${M}(x,y,t)$, whose complex conjugate we denote as $M^*$. To leading order in $\delta$, the wavy displacement $\eta(x,y,t)$ of the interface and the vertical velocity $w_s(x,y,t)$ at the interface are readily retrieved from ${M}$ as:
 \begin{align}
\eta(x,y,t) & = \frac{1}{\sqrt{2}} {M}(x,y,t)e^{-it} + \text{c.c.} \, , \label{eq:etavsM}\\
w_s(x,y,t) & = -\frac{i}{\sqrt{2}} {M}(x,y,t) e^{-it} + \text{c.c.}   \, .  \label{eq:wvsM}
 \end{align}

The reduced model consists of the following coupled evolution equations for the demodulated complex wave amplitude ${M}$ and the standard Eulerian velocity field ${\bf U}$ of the background flow: 
\begin{align}
& \partial_t {M} + \bU \cdot \bnabla M + \frac{1}{2}(\bnabla \cdot \bU)M-\frac{i}{4} (\Delta {M} + {M}) +i {M}^* f_1(D)\{ {M}^2\} +i {M} f_2(D)\{ |M|^2\} = 0 \, , \label{eq:model1}\\
& \partial_t {\bf U} + (\bnabla \times {\bf U})\times({\bf U+{\bf u}^s})  = -\bnabla P \, , \qquad \bnabla \cdot {\bf U}=0 \, . \label{eq:model2}
\end{align}
The expressions of the effective horizontal flow ${\bU}$ entering the wave equation and of the Stokes drift velocity ${\bf u}^s$ are:
\begin{align}
 \bU (x,y,t) & = \left( 1+ \frac{\Delta}{4}  \right) \left\{  \int_{-\infty}^0 2 e^{2z} {\bf U}_\perp \, \mathrm{d}z \right\}  \, ,  \label{eq:defbU} \\ 
 {\bf u}^s ({\bf x},t)  & =  2 e^{2z} \left[ \bP -\frac{1}{4} \bnabla \times (\bnabla \times \bP) - \frac{1}{2} (\bnabla \cdot \bP) {\bf e}_z \right] \, , \label{eq:usVY} 
\end{align}
where the subscript $\perp$ refers to the horizontal components only and
\begin{align}
\bP(x,y,t)=\frac{i}{2} \left( {M} \bnabla {M}^* - {M}^* \bnabla {M}\right)   \, .
\end{align}
The governing equation~(\ref{eq:model2}) for the background flow is the Craik-Leibovich equation, where the Stokes drift ${\bf u}^s$ is cast under the solenoidal form~(\ref{eq:usVY}) introduced by \citet{vanneste2022stokes}. That is, in the present formulation the Eulerian velocity ${\bf U}$ of the background flow,  the Stokes drift ${\bf u}^s$ and the Lagrangian flow ${\bf U}+{\bf u}^s$ are all divergence-free.  The Stokes drift $ {\bf u}^s$ features a nonzero vertical component at the free surface, ${\bf u}^s|_0 \cdot {\bf e}_z=- \bnabla \cdot \bP$, where the subscript $|_0$ means `evaluated at $z=0$'. In line with~\citet{andrews1978exact}, we impose an impenetrable boundary condition for the Lagrangian flow ${\bf U}+{\bf u}^s$ at $z=0$ (see also \citet{vanneste2026consistent,onuki2026reduced}), which corresponds to the following boundary condition for the standard Eulerian flow:
\begin{align}
{\bf U} |_0 \cdot {\bf e}_z = - {\bf u}^s|_0 \cdot {\bf e}_z = \bnabla \cdot \bP \, . \label{eq:surfaceBC}
\end{align}
For subsequent analysis as well as for practical implementation of the Stokes drift~(\ref{eq:usVY}) in a numerical solver, we note that the curl and the divergence of $\bP$ take the relatively compact forms:
\begin{align}
\bnabla \cdot \bP & =  \frac{i}{2} \left( {M} \Delta {M}^* - {M}^* \Delta {M}\right)   \, , \\
\bnabla \times \bP & =  iJ(M,M^*) {\bf e}_z \, , \label{eq:curlP}
\end{align}
where $J(f,g)=\partial_x(f) \partial_y(g)-\partial_y(f) \partial_x(g)$ denotes the Jacobian.


Turning to the wave equation, the first four terms in~(\ref{eq:model1}) correspond to the reduced wave equation derived in part I, see equation (8.27) in~\citet{galletpart1}. These terms describe the slow modulation of the complex wave amplitude as a result of  propagation at the group velocity and advection by the background flow. As compared to the reduced wave equation derived in part I, equation~(\ref{eq:model1}) contains two additional, cubic terms that encode nonlinear wave-wave interactions. The operator $D$ corresponds to multiplication by the horizontal wavenumber $k$ in spectral space, and therefore the operators $f_1(D)$ and $f_2(D)$ entering the cubic terms correspond to multiplication by $f_1(k)$ and $f_2(k)$ in spectral space, where the functions $f_1$ and $f_2$ are:
\begin{align}
f_1(D) & =\frac{D^5+12D^4-40D^3+32D^2+80D-64}{32(4-D)} \, , \label{eq:deff1} \\
f_2(D) & =- \left(\frac{D^2}{4}-1 \right)^2 \, . \label{eq:deff2} 
\end{align}
The relatively complex form of the nonlinear terms stems from the intricate four-wave interaction kernel for surface gravity waves~\citep{krasitskii1994reduced}. Because $M$ is narrow-banded, $M^2$ spans dimensionless wavenumbers $k \lesssim 2$ only and the vanishing denominator of~(\ref{eq:deff1}) for wavenumber $k=4$ is not an issue when computing $f_1(D)\{ {M}^2\}$. 



\section{Conservation laws\label{sec:conservation}}

We now discuss the invariants of the coupled model, focusing on the case of a doubly periodic domain in the horizontal directions. 

\subsection{Wave action}

As for the one-way coupled model, the two-way coupled model conserves wave action. Multiply equation~(\ref{eq:model1}) with ${M}^*$ before adding the complex conjugate and space-averaging to obtain:
\begin{align}
\frac{\mathrm{d}}{\mathrm{d} t} \la |{M}|^2 \ra & = 0 \, ,  \label{eq:actioncons}
\end{align}
where $\la \cdot \ra$ denotes a horizontal area average over $x$ and $y$. To obtain equation~(\ref{eq:actioncons}) and in the following, we make extensive use of the formula:
\begin{align}
\la a {\cal F}(D) \{ b \} \ra= \la  b {\cal F}(D) \{ a \} \ra \, , \label{eq:Dswitch}
\end{align}
valid for any function ${\cal F}$ and two fields $a$ and $b$, as easily deduced by decomposing both fields $a$ and $b$ into spatial Fourier series.


\subsection{Total energy}

Under the proposed form the coupled system also conserves  a mechanical-energy invariant. In part I we identified a wave-energy invariant that is conserved when the background flow is steady. The mechanical-energy invariant of the coupled model consists of both wave energy and kinetic energy from the background flow. To establish the conservation of this global invariant, it proves useful to introduce the following notations for vertical integration and weighted vertical integration:
\begin{align}
\overline{\cdot}= \int_{-\infty}^0 (\cdot) \mathrm{d}z \, ,  \qquad \hat{\cdot}= 2 \overline{e^{2z} (\cdot)} = \int_{-\infty}^0 2 e^{2z} (\cdot) \mathrm{d}z  \, .
\end{align}

The wave energy evolution equation is obtained by multiplying equation~(\ref{eq:model1}) with $\partial_t M^*$ before space averaging, subtracting the complex conjugate and dividing by $i$~\citep{danioux2015concentration}. After various integrations by parts and applications of formula~(\ref{eq:Dswitch}) we obtain:


\begin{align}
\frac{\mathrm{d}}{\mathrm{d} t}  \la \bP \cdot \bU + \frac{1}{4} \left( |\bnabla_\perp {M}|^2- |{M}|^2\right)  + \frac{1}{2} M^{*2} f_1(D) \{ M^2 \}+ \frac{1}{2} |M|^2 f_2(D) \{ |M|^2 \} \ra & = \la \bP \cdot  \partial_t {\bU} \ra\, , \label{eq:dTwaveE}
\end{align}

To obtain an evolution equation for the kinetic energy of the background flow, we take the dot-product of equation~(\ref{eq:model2}) with the divergence-free Lagrangian velocity field ${\bf U}+{\bf u}^s$. After horizontal average and vertical integration using the boundary condition~(\ref{eq:surfaceBC}) we simply obtain:
\begin{align}
\frac{\mathrm{d}}{\mathrm{d} t} \la \frac{\overline{{\bf U}^2}}{2} \ra & = -  \la \overline{{\bf u}^s \cdot \partial_t {\bf U} }\ra\, .
\end{align}
Inserting expression~(\ref{eq:usVY}) for the Stokes drift yields:
\begin{align}
\nonumber \frac{\mathrm{d}}{\mathrm{d} t} \la \frac{\overline{{\bf U}^2}}{2} \ra & =   - \la \partial_t \hat{\bf U} \cdot \left[  \bP -\frac{1}{4} \bnabla \times (\bnabla \times \bP) - \frac{1}{2} (\bnabla \cdot \bP) {\bf e}_z  \right] \ra \\
& = - \la \partial_t \hat{\bf U}_\perp \cdot \left[  \bP -\frac{1}{4} \bnabla \times (\bnabla \times \bP)  \right] - \frac{1}{2} \partial_t (\hat{W}) \bnabla \cdot \bP \ra \, , \label{eq:dtU2temp}
\end{align}
where $W$ denotes the vertical component of ${\bf U}$. Applying the $\hat{\cdot}$ operator to $\bnabla \cdot {\bf U} = 0$ gives:
\begin{align}
\hat{W} & = W|_0 + \frac{1}{2} \bnabla_\perp \cdot \hat{\bf U}_ \perp =  \bnabla_\perp \cdot \bP  + \frac{1}{2} \bnabla_\perp \cdot \hat{\bf U}_ \perp \, . \label{eq:What}
\end{align}
Substituting into~(\ref{eq:dtU2temp}) and performing a few integrations by parts leads to:
\begin{align}
\frac{\mathrm{d}}{\mathrm{d} t} \la \frac{\overline{{\bf U}^2}}{2} \ra & = \frac{1}{4} \frac{\mathrm{d}}{\mathrm{d} t} \la (\bnabla \cdot \bP)^2 \ra - \la \bP \cdot \partial_t \bU \ra \, . \label{eq:dTflowE}
\end{align}
Adding equations~(\ref{eq:dTwaveE}) and~(\ref{eq:dTflowE}) leads to conservation of the mechanical-energy invariant $E$ under the form $\frac{\mathrm{d} E}{\mathrm{d} t}  =  0$, where:
\begin{align}
& E  = \la \frac{\overline{{\bf U}^2}}{2} + \bP \cdot \bU +  \frac{1}{4} \left( |\bnabla {M}|^2- |{M}|^2\right)  + \frac{1}{2} M^{*2} f_1(D) \{ M^2 \}+ \frac{1}{2} |M|^2 f_2(D) \{ |M|^2 \} - \frac{1}{4}  (\bnabla \cdot \bP)^2   \ra  \, . \label{eq:energyE}
\end{align}

As a side note, one may want to express this energy invariant in terms of the kinetic energy of the Lagrangian mean flow ${\bf U}+{\bf u}^s$. Through integration by parts and using expression~(\ref{eq:usVY}) for the Stokes drift one can show that:
\begin{align}
\la \overline{{\bf U} \cdot {\bf u}^s} \ra & =  \la \bP \cdot \bU \ra - \frac{1}{2} \la (\bnabla \cdot \bP)^2 \ra \, .
\end{align}
We isolate $\la \bP \cdot \bU \ra$ from this expression and substitute into~(\ref{eq:energyE}) to obtain: 
\begin{align}
E & = \la \frac{\overline{({\bf U}+ {\bf u}^s)^2}}{2} \ra  \\
\nonumber & + \la \frac{1}{4} \left( |\bnabla {M}|^2- |{M}|^2\right)  +  \frac{1}{2} M^{*2} f_1(D) \{ M^2 \}+ \frac{1}{2} |M|^2 f_2(D) \{ |M|^2 \} + \frac{1}{4}  (\bnabla \cdot \bP)^2  - \frac{\overline{( {\bf u}^s)^2}}{2} \ra \, .
\end{align}
The energy invariant $E$ thus consists of the kinetic energy of the Lagrangian mean flow ${\bf U}+{\bf u}^s$, plus terms involving the wave amplitude $M$ only.

\subsection{Horizontal momentum\label{sec:horizmomentum}}

The coupled model conserves total horizontal momentum provided we exclude any background pressure gradient (as well as any other external force). We thus assume that the pressure variable $P$ in equation~(\ref{eq:model2}) satisfies the periodic boundary conditions in the horizontal directions. Applying horizontal average and vertical integration to the horizontal components of equation~(\ref{eq:model2}), using again the subscript $\perp$ to denote the horizontal components, we obtain:
\begin{align}
\nonumber \frac{\mathrm{d}}{\mathrm{d} t}  \la \overline{{\bf U}_\perp} \ra &  = - \la \overline{ (\bnabla \times {\bf U})\times {\bf U} } \ra_\perp - \la \overline{  (\bnabla \times {\bf U})\times {\bf u}^s } \ra_\perp \\
\nonumber & = -  \la  \overline{ ( {\bf U} \cdot \bnabla) {\bf U}_\perp } \ra -  \la W|_0 {{\bf u}^s}_\perp|_0 \ra - \la ({\bf u}^s|_0 \cdot {\bf e}_z) {\bf U}_\perp|_0 \ra - \la \overline{  {\bf U} \times (\bnabla \times {\bf u}^s)  } \ra_\perp \\
\nonumber & = -  \la (W|_0+{\bf u}^s|_0 \cdot {\bf e}_z) {\bf U}_\perp|_0 \ra -  \la W|_0 {{\bf u}^s}_\perp|_0 \ra - \la ({\bf u}^s|_0 \cdot {\bf e}_z) {\bf U}_\perp|_0 \ra - \la \overline{  {\bf U} \times (\bnabla \times {\bf u}^s)  } \ra_\perp \\
 & = -  2 \la (\bnabla \cdot \bP) \left( 1+ \frac{\Delta}{4}\right) \{ \bP \} \ra - \la \overline{  {\bf U} \times (\bnabla \times {\bf u}^s)  } \ra_\perp \, , \label{eq:tempdtUperp}
\end{align}
where we have used the surface boundary condition~(\ref{eq:surfaceBC}) to obtain the last line. The curl of the Stokes drift can be written as:
\begin{align}
 \bnabla \times {\bf u}^s = 2 e^{2z} \left( 1+ \frac{\Delta}{4}\right) \left\{ - 2 \bP \times {\bf e}_z +  \bnabla \times \bP  \right\} \, ,
\end{align}
which leads to:
\begin{align}
 \la \overline{  {\bf U} \times (\bnabla \times {\bf u}^s)  } \ra_\perp & = \la -2 \hat{W}  \left( 1+ \frac{\Delta}{4}\right) \{ \bP \} \ra + \la \bU \times (\bnabla \times \bP) \ra \\
 & = -  2 \la (\bnabla \cdot \bP) \left( 1+ \frac{\Delta}{4}\right) \{ \bP \} \ra - \la (\bnabla \cdot \bU) \bP \ra + \la \bU \times (\bnabla \times \bP) \ra  \, ,
\end{align}
where we have inserted~(\ref{eq:What}) to obtain the last equality. Substituting into~(\ref{eq:tempdtUperp}) finally leads to:
\begin{align}
\frac{\mathrm{d}}{\mathrm{d} t}  \la \overline{{\bf U}_\perp} \ra &  =   \la (\bnabla \cdot \bU) \bP \ra - \la \bU \times (\bnabla \times \bP) \ra  \, . \label{eq:dtUperp}
\end{align}
Secondly, we evaluate the time derivative of $\la {\bf u}^s \ra = \la {\bP} \ra$. The contributions from the nonlinear and dispersive terms of equation~(\ref{eq:model1}) vanish and we are left with:
\begin{align}
\nonumber \frac{\mathrm{d}}{\mathrm{d} t}  \la \bP \ra &  =  \frac{i}{2} M \bnabla \left( -\bU \cdot \bnabla M^* - \frac{1}{2}(\bnabla \cdot \bU) M^*  \right) +  \frac{i}{2} (\bnabla M^*) \cdot  \left( -\bU \cdot \bnabla M - \frac{1}{2}(\bnabla \cdot \bU) M  \right) + \text{c.c.} \\
\nonumber & = - \la (\bnabla \cdot \bU) \bP \ra  + i \left[ \la (\bU \cdot \bnabla M^*) \bnabla M \ra -  \la (\bU \cdot \bnabla M) \bnabla M^* \ra \right] \\
\nonumber & = - \la (\bnabla \cdot \bU) \bP \ra + \la \bU \times [ i J(M,M^*) {\bf e}_z ]   \ra \\
& = - \la (\bnabla \cdot \bU) \bP \ra + \la \bU \times (\bnabla \times \bP)   \ra \, , \label{eq:dtcalP}
\end{align}
where we have inserted~(\ref{eq:curlP}) to obtain the last equality. Adding equations~(\ref{eq:dtUperp}) and~(\ref{eq:dtcalP}) finally leads to momentum conservation under the form:
\begin{align}
\nonumber \frac{\mathrm{d}}{\mathrm{d} t}  \la \overline{{\bf U}_\perp + {\bf u}^s} \ra & = \nonumber \frac{\mathrm{d}}{\mathrm{d} t}  \la \overline{{\bf U}_\perp} + \bP \ra = {\bf 0} \, .
\end{align}

\section{Derivation of the Craik-Leibovich equation\label{sec:derivation}}

The asymptotic expansion leading to the coupled model consists of the expansion in part I leading to the reduced equation for the wave evolution, intertwined with the standard expansion leading to the CL equation. The present section focuses on obtaining the CL equation, the modifications to the reduced wave equation being the topic of section~\ref{sec:derivationNL}.

\subsection{Governing equations}

We consider narrow-band surface gravity waves around a central frequency $\omega_0$, with a small, ${\cal O}(\delta^2)$ relative bandwidth. As in part I, the flow is incompressible and we non-dimensionalize the equations using $g$ and $\omega_0$. The dimensionless equations read:
\begin{align}
\partial_t \eta + {\bf u}_{\perp}|_\eta \cdot \bnabla_\perp \eta & = w|_\eta \, , \label{eq:eta}\\
p_\eta &  = \eta \, , \\
\Delta p & = - \bnabla \cdot [({\bf u} \cdot \bnabla) {\bf u}] \, , \\
\partial_t w|_0 + ({\bf u} \cdot \nabla w)|_0 & = -(\partial_z p)|_0 \, ,  \label{eq:w}
\end{align}
where $p$ denotes the departure from the hydrostatic pressure profile of the motionless state, divided by density. 
To describe the evolution of the background flow, we complement this set of equations with the full vorticity equation:
\begin{align}
\partial_t (\bnabla \times {\bf u}) & = \bnabla \times [{\bf u} \times (\bnabla \times {\bf u})] \, , \label{eq:Omega}
\end{align}
together with the incompressibility constraint $\bnabla \cdot {\bf u}=0$. The coupled system of reduced equations is obtained by considering waves of  ${\cal O}(\delta)$ amplitude interacting with a background flow whose dimensionless speed is ${\cal O}(\delta^2)$.

We consider the equations above in a domain $[0, L]\times[0, L] \times (-\infty,0]$ with periodic boundary conditions in the horizontal directions, and we seek a solution under the form:
\begin{align}
{\eta} & =\delta \eta_1({\bf x},t,T) + \delta^2  \eta_2({\bf x},t,T) + \dots \, , \label{eq:expansion}\\
{p} & =\delta p_1({\bf x},t,T) + \delta^2  p_2({\bf x},t,T) + \dots  \, , \\
{{\bf u}} & =\delta {\bf u}_1({\bf x},t,T) + \delta^2  {\bf u}_2({\bf x},t,T) + \dots \, ,
\end{align}
where the slow time variable is $T=\delta^2 t$.  Finally, we denote as  $\la{\cdot} \ra_t$ an average with respect to the fast time variable $t$.

\subsection{${\cal O}(\delta)$: irrotational waves}

To order $\delta$, equations~(\ref{eq:eta}-\ref{eq:Omega}) reduce to:
\begin{align}
{\cal L}\{ {\bf X}_1\} & =0 \, , \\
\partial_t (\bnabla \times {\bf u}_1) & ={\bf 0} \, , \label{eq:Omega1} 
\end{align}
where ${\bf X}_i=(\eta_i, p_i, {\bf u}_i)^T$ and we have defined the linear operator ${\cal L}$ such that:
\begin{align}
{\cal L}\{ {\bf X}_i\} & = \begin{pmatrix} \partial_t \eta_i - w_i|_0 \\ \eta_i-p_i|_0 \\ \Delta p_i \\ \partial_t w_i|_0 + \partial_z p_i|_0 \end{pmatrix}
\end{align}
We consider the same lowest-order solution as in part I: irrotational waves around unit dimensionless angular frequency. We denote the corresponding velocity field as ${\bf u}_1 = \tilde{\bf u}_1({\bf x},t,T)$ where the tilde indicates a `wavy' flow that vanishes under fast-time average. That is, $\la \tilde{\bf u}_1 \ra_t={\bf 0}$ and the same holds in the following for any quantity with a tilde. Such irrotational waves trivially satisfy the lowest-order vorticity equation~(\ref{eq:Omega1}) since $\bnabla \times \tilde{\bf u}_1={\bf 0}$.

\subsection{${\cal O}(\delta^2)$: bound irrotational waves + slow background flow}

To order $\delta^2$, equations~(\ref{eq:eta}-\ref{eq:Omega}) lead to:
\begin{align}
{\cal L}\{ {\bf X}_2\} & = \text{NL}({\bf X}_1,{\bf X}_1) \, , \label{eq:wavy2}  \\
\partial_t (\bnabla \times {\bf u}_2) & ={\bf 0} \, , \label{eq:Omega2} 
\end{align}
where NL denotes a generic quadratic nonlinear (bilinear in the following) term in the ${\bf X}_1$ fields. Because there are no three-wave interactions for surface gravity waves, such terms constitute a non-resonant forcing of the operator ${\cal L}$, i.e., they do not lead to any secular term. A particular solution to equations~(\ref{eq:wavy2}-\ref{eq:Omega2}) consists of irrotational bound waves for ${\bf X}_2$, that is, irrotational wavy motion $\tilde{\bf u}_2({\bf x},t,T)$ that does not obey the dispersion relation. In line with the standard Craik-Leibovich ordering, we also include the background flow in the ${\cal O}(\delta^2)$ solution: a slowly evolving background flow ${\bf U}_2({\bf x},T)$ is a solution to the homogeneous versions of equations (\ref{eq:wavy2}-\ref{eq:Omega2}). The surface deformation associated with such a background flow is ${\cal O}(\delta^4)$ and therefore we can safely neglect it, see Part I. The flow at order $\delta^2$ finally reads:
\begin{align}
{\bf u}_2({\bf x},t,T)= \tilde{\bf u}_2({\bf x},t,T) + {\bf U}_2({\bf x},T) \, .
\end{align}
We denote as $\bOmega({\bf x},T)=\bnabla \times {\bf U}_2$ the vorticity of the background flow. 

\subsection{${\cal O}(\delta^3)$: reduced equation for the leading-order waves + weak rotational waves}

To order $\delta^3$, equations~(\ref{eq:eta}-\ref{eq:Omega}) lead to:
\begin{align}
{\cal L}\{ {\bf X}_3\} & =  \begin{pmatrix} - \partial_T \eta_1 - {\bf U}_2 \cdot \bnabla \eta_1 \\ 0 \\  - \bnabla \cdot [({{\bf u}_1} \cdot \bnabla) {\bf U}_2 + ({\bf U}_2 \cdot \bnabla) {{\bf u}_1}] \\  - \partial_T w_1|_0 - {\bf U}_2 \cdot \bnabla w_1|_0  \end{pmatrix} + \text{NL}(\tilde{\bf X}_2,{\bf X}_1)  \, , \label{eq:wavy3}  \\
\partial_t (\bnabla \times {\bf u}_3) & = \bnabla \times ({\bf u}_1 \times \bOmega )  \, , \label{eq:Omega3}
\end{align}
where $\tilde{\bf X}_2$ refers to the wavy part of ${\bf X}_2$, with vanishing fast-time average. A solvability condition is obtained by demanding that the resonant forcing on the rhs of equation~(\ref{eq:wavy3}) vanish. In part I we considered infinitesimal waves: the nonlinear term $\text{NL}(\tilde{\bf X}_2,{\bf X}_1)$ was absent, and the column vector on the rhs was denoted as ${\bf R}=(R_1, R_2, R_3, R_4)$. Instead of solving the solvability condition, we derived a compact `reduced' evolution equation for the wave field, demanding that solutions to the reduced equation share the same leading-order solution and the same solvability condition as the original system. The resulting reduced equation corresponds to the linear part of equation~(\ref{eq:model1}). In contrast with part I, here we consider waves with finite, ${\cal O}(\delta)={\cal O}(\sqrt{\epsilon})$ amplitude. As a result, the new term $\text{NL}(\tilde{\bf X}_2,{\bf X}_1)$ arises on the rhs of equation~(\ref{eq:wavy3}). This term contributes nonlinear terms to the solvability condition, and thus to the reduced equation. We postpone the derivation of these nonlinear terms to section~\ref{sec:derivationNL}, and for now we simply write the reduced wave equation as:
\begin{align}
& \partial_t {M} + \bU \cdot \bnabla M + \frac{1}{2}(\bnabla \cdot \bU)M -\frac{i}{4} (\Delta {M} + {M}) + \text{nonlinear terms} = 0 \, . \label{eq:MnoNL}
\end{align}

%
%
%

Following again the standard derivation of the CL equation, we now consider the vorticity equation~(\ref{eq:Omega3}), which we integrate with respect to the fast time variable to obtain:
\begin{align}
\bnabla \times \tilde{\bf u}_3  = \bnabla \times ({\bxi}_1 \times \bOmega )  \, , \label{eq:u3}
\end{align}
where we have introduced the wavy displacement ${\bxi}_1({\bf x},t,T)$ defined by $\partial_t {\bxi}_1={\bf u}_1$. Equation~(\ref{eq:u3}) shows that the waves have weak vorticity as a result of their interaction with the background flow.

\subsection{${\cal O}(\delta^4)$: Craik-Leibovich equation for the background-flow evolution}

The last step of the CL expansion is to write the vorticity equation at ${\cal O}(\delta^4)$:
\begin{align}
\partial_t (\bnabla \times {\bf u}_4)+\partial_T(\bnabla \times {\bf u}_2) & = \bnabla \times [{\bf u}_1 \times (\bnabla \times {\bf u}_3) + {\bf u}_2 \times (\bnabla \times {\bf u}_2)]  \, .
\end{align}
The evolution equation for the background flow is obtained after fast-time averaging:
\begin{align}
\partial_T \bOmega & = \bnabla \times ( {\bf U}_2\times \bOmega) + \bnabla \times  \la{\bf u}_1 \times (\bnabla \times {\bf u}_3) \ra_t  \, . \label{eq:Omegatemp}
\end{align}
A technical step in CL's derivation~\citep{craik1976rational}  is to show that $\bnabla \times  \la{\bf u}_1 \times (\bnabla \times {\bf u}_3) \ra_t= \bnabla \times ( {\bf u}^s \times \bOmega)$, and after substitution we finally obtain the celebrated CL equation:
\begin{align}
\partial_T \bOmega & = \bnabla \times [ ({\bf U}_2 +{\bf u}^s)   \times \bOmega ]  \, . \label{eq:CL}
\end{align}
Uncurling this equation leads to equation~(\ref{eq:model2}) of the reduced model, which is cast in terms of the standard non-expanded variables.

Additionally, the present narrow-band wave field can feature significant modulations over a large, ${\cal O}(\delta^{-2})$ spatial scale only. For such slowly modulated waves, \citet{vanneste2022stokes} show that the leading-order Stokes drift takes the divergence-free form:
\begin{align}
{\bf u}^s & = \frac{1}{2} \bnabla \times \la  {\bf u}_1 \times \bxi_1 \ra_t \, . \label{eq:defus}
\end{align}
The surface deformation~(\ref{eq:etavsM}) and the surface vertical velocity~(\ref{eq:wvsM}) are associated with the following expressions for the wavy potential flow ${\bf u}_1=\tilde{{\bf u}}_1$ and the wavy displacement $\bxi_1$:
\begin{align}
{\bf u}_1 = - \frac{i}{\sqrt{2}} e^z e^{-it} \begin{pmatrix}
\partial_x M \\
\partial_y M \\
M
\end{pmatrix}
+ \text{c.c.} \, ,  \qquad \bxi_1 = \frac{1}{\sqrt{2}} e^z e^{-it} \begin{pmatrix}
\partial_x M \\
\partial_y M \\
M
\end{pmatrix} 
+ \text{c.c.} \, \, .
\end{align}
Using the relation~(\ref{eq:curlP}) these expressions lead to:
\begin{align}
 \frac{1}{2}  \times \la  {\bf u}_1 \times \bxi_1 \ra_t  & = e^{2z} \left[ \bP \times {\bf e}_z -\frac{i}{2}J(M,M^*) {\bf e}_z \right] = e^{2z} \left( \bP \times {\bf e}_z -\frac{1}{2} \bnabla \times \bP \right) \, ,
\end{align}
whose curl is expression~(\ref{eq:usVY}) for the Stokes drift.

\section{Derivation of the wave-wave interaction terms \label{sec:derivationNL}}

\subsection{The shortcut: crossing wave trains and stationary waves\label{sec:crossing}}

\begin{figure}
    \centerline{\includegraphics[width=10 cm]{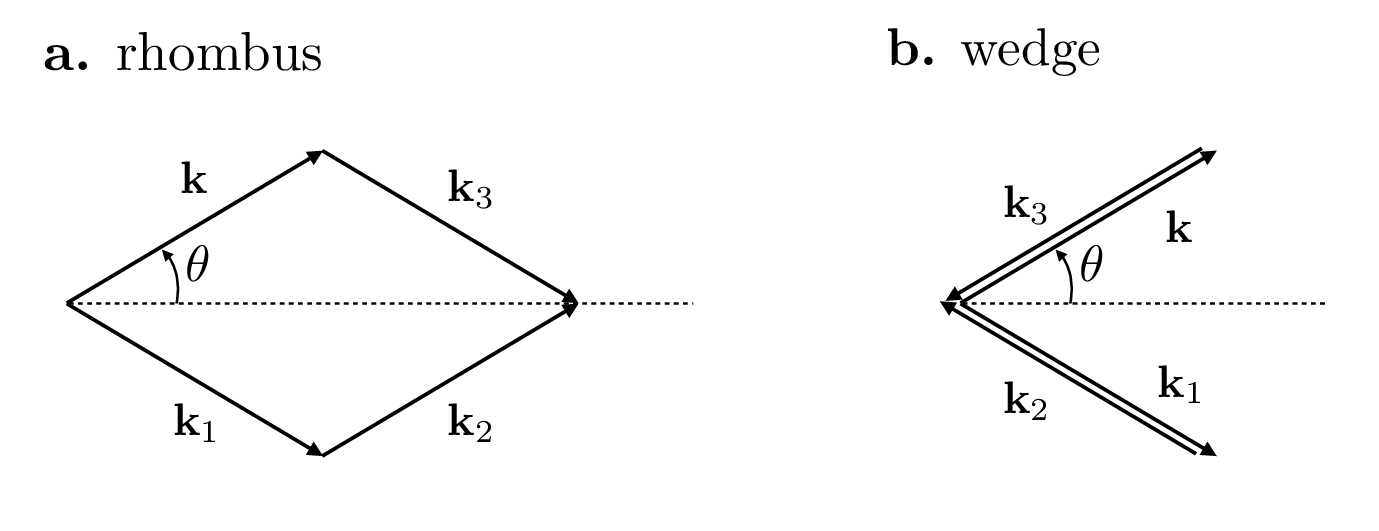} }
   \caption{\label{fig:quartet} Resonant quartets of wave modes with near-unit wavenumbers are organized either in a rhombus geometry (a), with effectively only two different wavevectors involved, or in a wedge geometry (b), with two pairs of opposite wavevectors.}
\end{figure}

A quick way to obtain the nonlinear terms of the reduced equation~(\ref{eq:model1}) begins by realizing that resonant quartets of wave modes with wavevectors ${\bf k}_1$, ${\bf k}_2$, ${\bf k}_3$ and ${\bf k}$ of near-unit norm are organized either in a rhombus geometry with ${\bf k}_1 \simeq {\bf k}_3$ and ${\bf k}_2 \simeq {\bf k}$ (permutations allowed) or in a wedge geometry with ${\bf k}_2 \simeq -{\bf k}_1$ and ${\bf k}_3 \simeq -{\bf k}$, see figure~\ref{fig:quartet}. There are effectively only two wavevectors involved in a rhombus quartet, corresponding to crossing wave trains, while there are two pairs of opposite wavevectors for the wedge geometry. For each quartet geometry the interaction coefficient can be deduced from the interaction kernel of the Zakharov equation. For the wedge geometry the interaction coefficient is given in~\citet{leblanc2024elementary}. For the rhombus geometry, the interaction coefficient is given in~\citet{onorato2006modulational,onorato2010freak} as part of their coupled envelope equations. 

We first validate the expression of the nonlinear terms entering the reduced equation~(\ref{eq:model1}) based on these studies. Then we explain how the nonlinear terms can easily be obtained by identification. The next subsections provide the longer version of the derivation, starting from the high-order-spectral equations governing weakly nonlinear waves.


Consider waves with wavevectors $\pm {\bf k}_a$ and $\pm {\bf k}_b$ only, where $| {\bf k}_a|=| {\bf k}_b|=1$. The complex amplitude $M$ is expanded as $M=\delta M_1(x,y,T) + \delta^3 M_3(x,y,t,T)$ and there is no background flow, $\bU={\bf 0}$. The lowest-order wave field reads:
\begin{align}
M_1 & =\frac{A_a(T)}{\sqrt{2}} e^{i {\bf k}_a \cdot {\bf x}} + \frac{A_{-a}(T)}{\sqrt{2}} e^{-i {\bf k}_a \cdot {\bf x}} +\frac{A_b(T)}{\sqrt{2}} e^{i {\bf k}_b \cdot {\bf x}} + \frac{A_{-b}(T)}{\sqrt{2}} e^{-i {\bf k}_b \cdot {\bf x}} \, . \label{eq:M1}
\end{align}
Without loss of generality, we choose the horizontal coordinates such that ${\bf k}_a=(\cos \theta, \sin \theta)$ and ${\bf k}_b=(\cos \theta, -\sin \theta)$ with $\cos \theta \geq 0$ (always feasible by appropriately choosing the direction of the $x$ axis, potentially swapping the two wavevectors). The lowest-order field~(\ref{eq:M1}) is indeed a solution to equation~(\ref{eq:model1}) at ${\cal O}(\delta)$. At ${\cal O}(\delta^3)$ we obtain:
\begin{align}
\partial_t { M_3} - \frac{i}{4} ( \Delta { M_3} + { M_3}) & = -\partial_T { M_1}  -i {M_1}^* f_1(D)\{ {M_1}^2\} -i {M_1} f_2(D)\{ |M_1|^2\}   \, . \label{eq:M3}
\end{align}
A solvability condition is obtained by demanding that the complex amplitude in front of $e^{i {\bf k}_a \cdot {\bf x}}$ on the rhs be zero. Collecting the various resonant terms, changing sign and multiplying by $\sqrt{2}$ leads to:
\begin{align}
\partial_T A_a & + \frac{i}{2}[f_1(2)+f_2(0)] |A_a|^2 A_a + \frac{i}{2} [2 f_1(0)+f_2(0)+f_2(2)] |A_{-a}|^2 A_a \label{eq:amplitudeeq}\\
\nonumber & + \frac{i}{2} [f_2(0)+f_2(|{\bf k}_a-{\bf k}_b|)+2f_1(|{\bf k}_a+{\bf k}_b|)] |A_{b}|^2 A_a \\
\nonumber & + \frac{i}{2} [f_2(0)+f_2(|{\bf k}_a+{\bf k}_b|)+2f_1(|{\bf k}_a-{\bf k}_b|)] |A_{-b}|^2 A_a \\
\nonumber & + \frac{i}{2} [2f_1(0)+f_2(|{\bf k}_a+{\bf k}_b|)+f_2(|{\bf k}_a-{\bf k}_b|)] A_b A_{-b} A_{-a}^* = 0 \, . 
\end{align}


Consider first crossing wave trains with nonzero $A_a$ and $A_b$ while $A_{-a}=A_{-b}=0$, corresponding to the wave field in~\citet{onorato2006modulational,onorato2010freak}.
Inserting expressions~(\ref{eq:deff1}-\ref{eq:deff2}) for the functions $f_1$ and $f_2$, the amplitude equation~(\ref{eq:amplitudeeq}) becomes:
\begin{align}
\partial_T A_a + \frac{i}{2} |A_a|^2 A_a + i \left( \frac{-\cos^5 \theta-2 \cos^4 \theta + 5 \cos^3 \theta -2 \cos^2 \theta - 3 \cos \theta + 2}{\cos \theta-2}\right) |A_b|^2 A_a & = 0 \, . \label{eq:SNLlike}
\end{align}
The nonlinear coupling terms in equation~(\ref{eq:SNLlike}) correspond precisely to those obtained by \citet{onorato2006modulational,onorato2010freak} (consider for instance equations (14) and (17) in~\citet{onorato2010freak}, setting their parameters to $k=\cos \theta$, $l=\sin \theta$, $\kappa=1$, $\omega(\kappa)=1$ and noticing that they include a factor of $2$ in their definition of the second coupling coefficient).

When we also include nonzero $A_{-a}$ and $A_{-b}$, additional terms arise in the solvability condition, see the rhs of~(\ref{eq:amplitudeeq}). Two of these terms correspond to the crossing-wave-train interactions between $A_a$ and waves with wavevector $-{\bf k}_a$ or $-{\bf k}_b$. They are associated with quartets of rhombus geometry. The very last term on the rhs of~(\ref{eq:amplitudeeq}), however, is different. It corresponds to a quartet interaction in the wedge geometry, involving four different wavevectors. Inserting the expressions of $f_1$ and $f_2$ this term reads:
\begin{align}
-\frac{i}{8} [7+ \cos(4 \theta)] A_b A_{-b} A_{-a}^*  \, ,
\end{align}
which corresponds precisely to the interaction coefficient reported in~\citet{leblanc2024elementary} (see the expression below his equation (C6)).

The quick way of obtaining the nonlinear term in the reduced equation~(\ref{eq:model1}) thus consists in assuming the phase-invariant form $i {M}^* f_1(D)\{ {M}^2\}+i {M} f_2(D)\{ |M|^2\}$ for this nonlinear term, with unknown functions $f_1(D)$ and $f_2(D)$. From this general form, one can derive the amplitude equation (\ref{eq:amplitudeeq}). Identifying the terms of this equation with those obtained by~\citet{onorato2010freak} and~\citet{leblanc2024elementary} readily provides the expressions of the functions $f_1(D)$ and $f_2(D)$.

\subsection{The scenic route: high-order spectral equations}

We now provide a more standard derivation of the nonlinear terms entering equation~(\ref{eq:MnoNL}), with the goal of obtaining equation~(\ref{eq:model1}). The reader convinced by the previous argument may be willing to skip ahead directly to section \ref{sec:viscous}.

A convenient aspect of multiple-time asymptotics is that, when properly scaled, the various physical effects arise {\it additively} in the solvability condition, and therefore in the reduced equation. In other words, the reduced equation consists of the sum of the wave-mean flow interaction terms computed for infinitesimal waves in part I, together with the nonlinear terms arising from four-wave interaction in the absence of background flow. To compute the latter, we use as starting point the standard high-order-spectral equations for surface gravity waves in the absence of background flow~\citep{dommermuth1987high,west1987new,onorato2002freely,dyachenko2004weak,zhang2022numerical,higgins2024numerical}. These equations couple the surface elevation $\eta(x,y,t)$ and the vertical velocity at the surface $w_s(x,y,t)$ through:
\begin{align}
\begin{split}\label{eq:HOS_eta}
    \partial_t \eta  - w_s ={}& -\bnabla \cdot \left( \eta \bnabla D^{-1} w_s \right) - D \{ \eta w_s \} + D\left\{ \eta D\{ \eta w_s\} \right\}  - \frac{1}{2}D \left\{ \eta^2 Dw_s \right\} + \frac{1}{2} \Delta(\eta^2 w_s) \, ,
\end{split}\\
\begin{split}\label{eq:HOS_w}
    \partial_t w_s  + D\eta ={}&  - \frac{1}{2} D \left\{ (\bnabla D^{-1} w_s)^2 - w_s^2 \right\}  - D \left\{ w_s  D\{\eta w_s\} \right\} + D \left\{ \eta w_s  D w_s\right\} \, . 
\end{split}
\end{align}
where nonlinearities have been retained up to cubic order only, which is sufficient to capture four-wave interactions. As in part I, we introduce the complex variable $\chi(x,y,t)=D^{-\frac{1}{4}} \eta / \sqrt{2}+ iD^{-\frac{3}{4}} w_s /\sqrt{2}$, or equivalently:
 \begin{align}
\eta & =\frac{1}{\sqrt{2}} (D^{{\frac{1}{4}}} \chi + D^{{\frac{1}{4}}} \chi^*) \, , \label{eq:etavschi}\\
w_s & =-\frac{i}{\sqrt{2}} (D^{{\frac{3}{4}}} \chi - D^{{\frac{3}{4}}} \chi^*) \, . \label{eq:wvschi}
 \end{align}
An evolution equation for $\chi$ is obtained from the linear combination of equations $(D^{-{\frac{1}{4}}}\{$(\ref{eq:HOS_eta})$\} + i D^{-{\frac{3}{4}}}\{$(\ref{eq:HOS_w})$\})/\sqrt{2}$:
\begin{align}
\partial_t \chi +i D^{\frac{1}{2}} \chi = {\cal R}_\text{quad} + {\cal R}_\text{cubic} \, , \label{eq:chi}
\end{align}
where ${\cal R}_\text{quad}$ (respectively ${\cal R}_\text{cubic}$) contains the quadratic (respectively cubic) terms in $\chi$, their expressions being given in appendix~\ref{sec:rhseqchi}. We seek a solution to equation~(\ref{eq:chi}) under the following multiple-time expansion:
\begin{align}
\chi = \delta \chi_1(x,y,t,T) + \delta^2 \chi_2(x,y,t,T) + \delta^3 \chi_3(x,y,t,T) + \dots \, . \label{eq:expchi}
\end{align}
To order $\delta$, equation~(\ref{eq:chi}) gives:
\begin{align}
\partial_t \chi_1 +i D^{\frac{1}{2}} \chi_1 = 0 \, ,
\end{align}
We consider a narrow-band solution around unit angular frequency under the form:
 \begin{align}
 \chi_1 & =   \left[ \sum_{\bf k} A_{\bf k}(T) e^{i {\bf k}\cdot {\bf x}} \right] e^{- it} \, , \label{eq:solchi1}
 \end{align}
 where the sum is over the wavevectors ${\bf k}$ compatible with the periodic boundary conditions in the horizontal directions, and the solution satisfies the following narrow-band property:
\begin{align}
\textit{narrow-band property: } & A_{\bf k} = {\cal O}(1) \text{ only for } {\bf k} \text{ such that } k=1 + {\cal O}(\delta^2)   \label{nbassump} \\
\nonumber & A_{\bf k} \simeq 0 \text{ otherwise}
\end{align}
 A consequence of the narrow-band property~(\ref{nbassump}) is that, for any $\alpha$, we have:
\begin{align}
D^{\alpha}{\chi_1} & ={\chi_1} [1+ {\cal O}(\delta^2)] \, . \label{eq:NBcalM}
\end{align}
To order $\delta^2$, equation~(\ref{eq:chi}) gives:
\begin{align}
\partial_t \chi_2 +i D^{\frac{1}{2}} \chi_2 = {\cal R}_\text{quad}|_{\chi_1} \, , \label{eq:chi2}
\end{align}
where the rhs corresponds to the term ${\cal R}_\text{quad}$ in which we substitute $\chi = \chi_1$ and simplify using $D^\alpha{\chi_1}=\chi_1$, valid to the desired order of accuracy. As discussed around equation~(\ref{eq:wavy2}), ${\cal R}_\text{quad}|_{\chi_1}$ contains no resonant terms because there are no three-wave resonances among deep-water surface gravity waves. The solution for $\chi_2$ can be written as:
\begin{align}
\chi_2 = \chi_2^{(-2)}(x,y,T) e^{-2it} + \chi_2^{(0)}(x,y,T)  + \chi_2^{(2)}(x,y,T) e^{2it} \, , \label{eq:formchi2}
\end{align}
where $\chi_2^{(-2)}$, $\chi_2^{(0)}$ and $\chi_2^{(2)}$ correspond to the complex amplitudes of the response at temporal harmonics $-2$, $0$ and $+2$, respectively. Inserting~(\ref{eq:formchi2}) into equation~(\ref{eq:chi2}) and isolating the different harmonics, we obtain:
\begin{align}
\chi_2^{(-2)} e^{-2it} & = \frac{1}{4 \sqrt{2}} (D^{\frac{1}{2}}-2)^{-1} \left\{ D^{\frac{1}{4}} \{ (\bnabla {\chi_1})^2 \} + (2 D^{\frac{3}{4}} - D^{\frac{7}{4}} - D^{\frac{1}{4}} ) \{ {\chi_1}^2 \}  \right\} \, , \label{eq:chi2minus2}\\
\chi_2^{(0)} & = \frac{1}{2 \sqrt{2}} D^{-\frac{1}{4}} \{ |{\chi_1}|^2 -  |\bnabla {\chi_1}|^2 \} \, , \\
\chi_2^{(2)} e^{2it}  & = \frac{1}{4 \sqrt{2}} (D^{\frac{1}{2}}+2)^{-1} \left\{ D^{\frac{1}{4}} \{ (\bnabla {\chi_1}^*)^2 \} + (- 2 D^{\frac{3}{4}} + D^{\frac{7}{4}} - D^{\frac{1}{4}} ) \{ {\chi_1}^{*2} \}  \right\} \, .\label{eq:chi2plus2}
\end{align}
To order $\delta^3$, equation~(\ref{eq:chi}) gives:
\begin{align}
\partial_t \chi_3 +i D^{\frac{1}{2}} \chi_3 = -\partial_T \chi_1 + {\cal R}_\text{quad}|_{{\chi_1} \leftrightarrow \chi_2} + {\cal R}_\text{cubic}|_{{\chi_1}} \, , \label{eq:chi3}
\end{align}
where ${\cal R}_\text{quad}|_{{\chi_1} \leftrightarrow \chi_2}$ denotes the cross terms between $\chi_1$ and $\chi_2$ (and their complex conjugates) arising from ${\cal R}_\text{quad}$, and ${\cal R}_\text{cubic}|_{{\chi_1}}$ denotes the cubic term ${\cal R}_\text{cubic}$ in which we substitute $\chi=\chi_1$. 
For any wavevector ${\bf k}$ of near-unit norm, a solvability condition is obtained by multiplying equation~(\ref{eq:chi3}) with $e^{i(-{\bf k}\cdot {\bf x} +t)}$ and averaging over $x$, $y$ and fast time $t$, which leads to:
\begin{align}
0 = \la \la  e^{i(-{\bf k}\cdot {\bf x} +t)} \left[ -\partial_T \chi_1 + {\cal R}_\text{quad}|_{{\chi_1} \leftrightarrow \chi_2} + {\cal R}_\text{cubic}|_{{\chi_1}} \right] \ra \ra_{t}  \, , \label{eq:SCth}
\end{align}
where $|{\bf k}|=1 + {\cal O}(\delta^2)$. In appendix~\ref{sec:collecting}, we show that the solvability conditions can be recast in terms of $\chi_1$ and $\chi_1^*$ as:
\begin{align}
0 =  & \la \la  e^{i(-{\bf k}\cdot {\bf x} +t)}  \left[  \partial_T {\chi_1} + {\cal N}\{\chi_1 \}  \right] \ra \ra_{t} \, ,  \label{eq:SCrecast}
\end{align}
 where
 \begin{align}
 {\cal N}\{\chi_1 \} & =  \frac{i}{16} \bnabla {\chi_1}^* \cdot \bnabla \left[ \frac{D(D^2+4D-8)}{4-D}  \{ {\chi_1}^2 \}  \right]  - i \chi_1 \left(\frac{D^2}{4}-1 \right)^2 \{ |\chi_1|^2 \} \\
\nonumber & - \frac{i}{4}  {\chi_1}^* \left( \frac{D^4-4 D^3+6D^2-2D^2+10D-8}{D-4}\right)  \{ {\chi_1}^2\}    \, .
\end{align}
These solvability conditions lead to a cumbersome set of coupled ODEs governing the slow-time evolution of the complex amplitudes $A_{\bf k}(T)$ in~(\ref{eq:solchi1}). Instead of solving such solvability conditions, in part I we introduced an equivalent-solvability-condition method, where we design an evolution equation for $\chi$ that shares the  same leading-order solution and the same solvability conditions as the original system. From the form of the solvability condition~(\ref{eq:SCrecast}), it is apparent that the following reduced equation for $\chi$:
\begin{align}
   \partial_t \chi+ i D^{\frac{1}{2}} \chi  +  {\cal N}\{\chi \} & = 0 \,  \label{eq:reducedchi}
\end{align}
shares the same leading-order solution~(\ref{eq:solchi1}) and the same solvability conditions~(\ref{eq:SCrecast}) as the original system under the expansion~(\ref{eq:expchi}). An appealing aspect of the reduced equation~(\ref{eq:reducedchi}) is that it is phase-invariant and it features no quadratic terms anymore. As in part I, we expand the dispersion relation as $D^{\frac{1}{2}} \chi = ({3} \chi -  \Delta \chi)/4 + {\cal O}(\delta^5)$, valid in the narrow-band regime, and we introduce the demodulated complex amplitude ${\cal M}=\chi e^{it}$. Equation~(\ref{eq:reducedchi}) becomes:
\begin{align}
\partial_t {\cal M} - \frac{i}{4} ( \Delta {\cal M} + {\cal M})  +  {\cal N}\{{\cal M} \} & = 0 \, . \label{eq:calM}
\end{align}
In the narrow-band regime, the second term is of order $\Delta {\cal M} + {\cal M} \sim \delta^2  {\cal M} \sim \delta^3$, and the last, cubic term is also ${\cal O}(\delta^3)$. Equation~(\ref{eq:calM}) thus effectively describes the evolution of ${\cal M}$ with the slow time variable $T=\delta^2 t$. The neglected terms in this long-time evolution equation are  ${\cal O}(\delta^5)$.

\subsection{Remodeling}

In a similar fashion to normal-form analysis~\citep{guckenheimer2013nonlinear}, the nonlinear terms can be simplified by performing a near-identity change of variable. Consider the field:
\begin{align}
M = {\cal M} + {\cal M}^* {\cal S}\{{\cal M} \} \, , \label{eq:Mchange}
\end{align}
where 
\begin{align}
 {\cal S}\{{\cal M} \}=- \frac{D(D^2+4D-8)}{8(4-D)}  \{ {\cal M}^2\}   \, . 
\end{align}
Equation~(\ref{eq:Mchange}) is a near-identity change of variable, with $M$ equal to ${\cal M}$ up to ${\cal O}(\delta^3)$ corrections. 

Using $\partial_t{\cal M} = {\cal O}(\delta^3)$, the unsteady term in equation~(\ref{eq:calM}) simply becomes:
\begin{align}
{\partial_t {\cal M}} & = \partial_t M + {\cal O}(\delta^5) \, ,  \label{eq:changedt}
\end{align}
and the nonlinear term becomes
\begin{align}
{\cal N} \{ {\cal M} \} & = {\cal N} \{ M\}  + {\cal O}(\delta^5)  \, . \label{eq:changeNL}
\end{align}
To evaluate the dispersive term we first compute:
\begin{align}
\nonumber \Delta M + M & =   \Delta {\cal M} + {\cal M}  + 2 \bnabla {\cal M}^* \cdot \bnabla {\cal S}\{{\cal M} \} + {\cal M}^* \Delta {\cal S}\{{\cal M} \}  + \underbrace{( \Delta {\cal M}^* + {\cal M}^*)}_{\sim \delta^3} \underbrace{{\cal S}\{{\cal M} \}}_{\sim \delta^2} \\
\nonumber & =  \Delta {\cal M} + {\cal M} + 2 \bnabla {\cal M}^* \cdot \bnabla {\cal S}\{{\cal M} \} + {\cal M}^* \Delta {\cal S}\{{\cal M} \} + {\cal O}(\delta^5) \, , \\
& =  \Delta {\cal M} + {\cal M} + 2 \bnabla { M}^* \cdot \bnabla {\cal S}\{{ M} \} + { M}^* \Delta {\cal S}\{{ M} \} + {\cal O}(\delta^5) \, .
\end{align}
Using this equality the dispersive term becomes:
\begin{align}
 - \frac{i}{4} ( \Delta {\cal M} + {\cal M}) &  = - \frac{i}{4} ( \Delta { M} + { M})  + \frac{i}{2} \bnabla { M}^* \cdot \bnabla {\cal S}\{{ M} \} + \frac{i}{4} { M}^* \Delta {\cal S}\{{ M} \}   + {\cal O}(\delta^5) \, .\label{eq:changedisp} 
\end{align}
Inserting~(\ref{eq:changedt}), (\ref{eq:changeNL}) and (\ref{eq:changedisp}) into equation~(\ref{eq:calM}) and substituting the expressions of ${\cal N} \{ M \}$ and ${\cal S}\{{ M} \} $, we obtain:
\begin{align}
\partial_t { M} - \frac{i}{4} ( \Delta { M} + { M})   +i {M}^* f_1(D)\{ {M}^2\} +i {M} f_2(D)\{ |M|^2\}   & = 0 \, , \label{eq:Mtemp}
\end{align}
where $f_1(D)$ and $f_2(D)$ are given by~(\ref{eq:deff1}-\ref{eq:deff2}) and we neglect the subdominant, ${\cal O}(\delta^5)$ terms.
This reduced equation governs the evolution of weakly nonlinear narrow-band waves. Once again, the various physical ingredients are scaled in such a way that they arise additively in the solvability condition, and therefore in the reduced equation. In the presence of a background flow, one simply needs to include the background-flow terms $\bU \cdot \bnabla M + \frac{1}{2}(\bnabla \cdot \bU)M$ derived in part I in equation~(\ref{eq:Mtemp}), which finally leads to the reduced equation~(\ref{eq:model1}).

\section{Viscous effects\label{sec:viscous}}

Including viscous effects is a necessary step before implementing the model in most numerical solvers. For models framed in terms of the Lagrangian velocity, one may wonder whether the viscous terms should apply to the Eulerian or to the Lagrangian velocity. Because the present model is cast in terms of the Eulerian velocity and is derived through a systematic expansion, including viscosity is relatively straightforward. Denoting as $\nu^{\text{(dim)}}$ the dimensional kinematic viscosity, the dimensionless viscosity is $\nu=\nu^{\text{(dim)}} \omega_0^3/g^2$. Assuming that $\nu$ scales as $\delta^2$, the impact of viscosity on the equations of the coupled model is three-fold:
\begin{itemize}
\item Firstly, viscosity leads to an additional damping term in the ${\cal O}(\delta^3)$ solvability condition and therefore in the reduced equation. The damping term is reported in Lamb's book, see~\citet{lamb1930hydrodynamics} page 544, article 301, equation (9). In the present reduced equation, it corresponds to a term $- 2 \nu \Delta_\perp M$ on the lhs of equation~(\ref{eq:model1}). Using the narrow-band approximation, we simplify this term as $- 2 \nu \Delta_\perp M=2 \nu M + {\cal O}(\delta^5)$, where we have used $\Delta_\perp M=-M + {\cal O}(\delta^3)$ together with $\nu={\cal O}(\delta^2)$.
\item Secondly, viscosity enters the CL equation through a standard term $\nu \Delta {\bf U}$ on the rhs of~(\ref{eq:model2}). This term is readily obtained by keeping a small viscosity $\nu={\cal O}(\delta^2)$ in the standard CL expansion.
\item Finally, arbitrarily small viscosity affects the boundary condition for the tangential mean flow at the surface, because of effective wave-induced stresses. This boundary condition is discussed by various authors~\citep{longuet1953mass,unluata1970mass,madsen1978mass,xu1994wave,fujiwara2020mutual,fujiwara2020wave}, a compact derivation being proposed in \citet{seshasayanan2019surface}. The resulting boundary condition for the tangential background flow is $\partial_z {\bf U}_\perp|_0=\partial_z {\bf u}^s |_0$.
\end{itemize}
With these three modifications, the viscous coupled model reads:
\begin{align}
& \partial_t {M} + \bU \cdot \bnabla M + \frac{1}{2}(\bnabla \cdot \bU)M-\frac{i}{4} (\Delta {M} + {M}) +i {M}^* f_1(D)\{ {M}^2\} +i {M} f_2(D)\{ |M|^2\}  + 2\nu M= 0 \, , \label{eq:model1viscous}\\
& \partial_t {\bf U} + (\bnabla \times {\bf U})\times({\bf U+{\bf u}^s})  = -\bnabla P + \nu \Delta {\bf U}\, , \qquad \bnabla \cdot {\bf U}=0 \, . \label{eq:model2viscous} \\
& {\bf U}|_0 \cdot {\bf e}_z = \bnabla \cdot \bP \, , \qquad \partial_z {\bf U}_\perp|_0=\partial_z {\bf u}^s |_0 \, .\label{eq:BCviscous}
\end{align}
Viscous effects are dissipative and the model~(\ref{eq:model1viscous}-\ref{eq:BCviscous}) does not conserve energy nor action. However, horizontal momentum should be conserved in the absence of external forces and indeed it is. Repeating the algebra in section~\ref{sec:horizmomentum} for the viscous model leads to the following contributions from the viscous terms:
\begin{align}
\nonumber \frac{\mathrm{d}}{\mathrm{d} t}  \la \overline{{\bf U}_\perp} \ra &  = \text{p.t.} + \nu \partial_z \la {\bf U}_\perp \ra|_0 \,  \\
\nonumber &  = \text{p.t.} + \nu \partial_z \la {\bf u}^s \ra|_0 \, \\
&  = \text{p.t.} + 4 \nu \la \bP \ra \, , \label{eq:dtUviscous}
\end{align}
together with:
\begin{align}
\nonumber \frac{\mathrm{d}}{\mathrm{d} t}  \la \bP \ra &  = \text{p.t.} + i \nu \la -2 M \bnabla M^* + 2 M^* \bnabla M \ra \, \\
& =  \text{p.t.} - 4 \nu \la \bP \ra \, , \label{eq:dtusviscous}
\end{align}
where `p.t.' denotes the `previous terms' computed in section~\ref{sec:horizmomentum}. Upon adding equations~(\ref{eq:dtUviscous}) and~(\ref{eq:dtusviscous}) both the previous terms and the new viscous terms cancel, leading to conservation of horizontal momentum:
\begin{align}
\nonumber \frac{\mathrm{d}}{\mathrm{d} t}  \la \overline{{\bf U}_\perp + {\bf u}^s} \ra & = \nonumber \frac{\mathrm{d}}{\mathrm{d} t}  \la \overline{{\bf U}_\perp} + \bP \ra = {\bf 0} \, .
\end{align}

\section{Discussion: the minimal model versus the exactly conservative model \label{sec:discussion}}

In part I, we showed that a background flow with a dimensionless scale of order unity only affects the wave field through the rotational part of the near-surface effective flow $\bU$. That is, through a Helmholtz decomposition ${\bU}=- \bnabla \times {\psi(x,y,t) {\bf e}_z}+\bnabla S(x,y,t)$, we showed that only the streamfunction $\psi$ affects the waves to leading order, with $S$ having only a subdominant effect. Accordingly, we proposed a change of variable to remove the terms involving $S$ in the reduced wave equation. 
 By contrast, to obtain exact conservation of energy and momentum in the coupled model, we kept the full expression of $\bU$ in the wave equation~(\ref{eq:model1}). Such exact conservation is not a mandatory feature of a reduced model, however, and an alternative, `minimal' model can be obtained by performing the same change of variable as in part I. This would  turn the wave equation into (see \citet{galletpart1}):
\begin{align}
& \partial_t {M} + J(\psi,M) -\frac{i}{4} (\Delta {M} + {M}) +i {M}^* f_1(D)\{ {M}^2\} +i {M} f_2(D)\{ |M|^2\} = 0 \, , \label{eq:model1psi} \\
& \text{where } \, \psi(x,y,t)  = \left( \Delta_\perp^{-1}+\frac{1}{4} \right) \left\{  \int_{-\infty}^0 2 e^{2z} (\bnabla \times {\bf U}) \cdot {\bf e}_z \, \mathrm{d}z \right\} \, .
\end{align}
A reduced model with fewer coupling terms is obtained by coupling this equation with the CL equation~(\ref{eq:model1}). The resulting model readily encodes the fact that the divergent part of the near-surface flow does not affect the wave field to leading order. Such a model conserves wave action, but it only approximately conserves energy and horizontal momentum. The latter become adiabatic invariants for the various scalings of poloidal flow we can think about: slow poloidal flows with  $S={\cal O}(\delta^2)$ evolving on a ${\cal O}(1)$ spatial scale with an ${\cal O}(\delta^{-2})$ turnover time ; poloidal flows with $S={\cal O}(1)$ extending over a large spatial scale $1/\delta^2$ and evolving over a long, ${\cal O}(\delta^{-4})$ turnover time ; poloidal return flows beneath surface wave packets, corresponding to $S={\cal O}(\delta^2)$ over a spatial scale $1/\delta^2$ and propagating at the fast ${\cal O}(1)$ group velocity. The latter point is also consistent with Dysthe's equation~\citep{dysthe1979note}, which shows that return flows have a subdominant effect on the wave amplitude in a nonlinear Schr\"odinger framework.

To summarize, there is a tension between the search for a `minimal' model that includes terms up to a given asymptotic order only, with as few terms as possible in the equations, and the search for a `fully conservative' model that features exact invariants, as opposed to adiabatic ones. Abiding by the principle of `first do no harm', we focused on the fully conservative model throughout the article.


\section{Conclusion\label{sec:conclusion}}

The two-way coupled model derived here offers a compact description of surface wave-mean flow interactions, in a form that is well-suited for analytical study and/or implementation in standard pseudo-spectral solvers~(e.g. \citet{miquel2021coral}). Such numerical studies would shed light on various nonlinear phenomena of interest. For instance, while weak waves are deflected by a vortex, it has been experimentally observed that such deflection is suppressed as the waves are made stronger, because strong waves distort the background vortex~\citep{humbert2017wave}. The present coupled model offers a useful framework to study this nonlinear regime of wave refraction. 

Beyond the feedback on the background flow, we also expect wave-wave interactions to impact wave scattering. Consider a thought experiment where a wave maker sends waves at a given frequency towards a patch of organized Taylor-Green flow, similar to the situation considered in part I. We saw that efficient scattering only arises when the wavevector ${\bf k}_1$ of the incoming waves and the wavevector ${\bf K}$ of the background flow satisfy the resonance condition $|{\bf k}_1+{\bf K}|=1$. Now as a consequence of wave-wave interactions, 
the wavenumber $k_1$ of the incoming waves crucially depends on the wave amplitude: for a single uniform wave train $M \propto e^{i {{\bf k}_1 \cdot {\bf x}}}$ in a region where $\bU={\bf 0}$, equation~(\ref{eq:model1}) gives the wavenumber $k_1=\sqrt{1-4 |M|^2} \simeq 1- 2|M|^2$.
Whether resonant scattering arises or not depends on the strength of the waves. We can envision a situation where weak waves would be strongly scattered while strong waves would propagate almost unaffected by the background flow, or vice-versa.

Beyond thought experiments and laboratory setups, we have not hidden our interest for oceanic Langmuir cells. The present model offers a way forward for studying the feedback of the cells onto the surface waves that generate them, thereby characterizing the equilibrated state of Langmuir circulations.
To some extent, the model offers physical insight even in the absence of a numerical solution. 
The simplest possible setup for the CL instability is arguably one where the wave field consists of a single monochromatic plane wave propagating along ${\bf e}_x$. The dimensionless wavevector of the incoming waves is then ${\bf k}_1={\bf e}_x$. Together with a pre-existing vertically sheared flow along ${\bf e}_x$, the waves induce Langmuir cells through the CL instability. The standard eigenmode of this instability takes the form of a flow that depends on $y$ and $z$ only~\citep{craik1977generation,leibovich1977convective}, with a horizontal wavevector ${\bf K}=\pm K {\bf e}_y$ for the effective background flow (asymptotic models also point to very slow $x$-dependence in the nonlinear regime, see~\citet{chini2008strongly,chini2009asymptotically,zhang2015dynamic}). An immediate consequence is that $|{\bf k}_1+{\bf K}|=|{\bf e}_x \pm K{\bf e}_y|>1$, meaning that there is no resonant scattering of the wave field by the Langmuir cells in this simple geometry (outside the scale-separation regime where $K \ll 1$). As a result, we do not expect any modification to the leading-order wave field induced by the background flow, and instead we only expect weak bound waves, which are neglected at the level of approximation of the model. Beyond such a simple geometry, however, more realistic wave fields should include some angular spread around the dominant direction ${\bf e}_x$, and part of the angular spectrum could be scattered by the Langmuir cells. Additionally, we expect the background flow to eventually evolve towards fully 3D `Langmuir turbulence'~\citep{mcwilliams1997langmuir,harcourt2008large,grant2009characteristics,belcher2012global}, which could then efficiently scatter the surface wave field.

\medskip
\noindent  \textbf{Acknowledgements:} Yohei Onuki is warmly acknowledged for thoroughly reading an early version of this manuscript, providing insightful comments and suggestions.
The author thanks Alexandre Tlili for insightful comments and discussions as well.

\medskip
\noindent  \textbf{Funding:} This research is supported by the European Research Council under grant agreement 101124590.

\medskip
\noindent \textbf{Competing interests:} The author declares none.

\appendix

\section{Algebraic details regarding the computation of nonlinear terms}

\subsection{Quadratic and cubic terms in the $\chi$ equation\label{sec:rhseqchi}}

The rhs of equation~(\ref{eq:chi}) features quadratic and cubic terms in $\chi$. The quadratic term reads:
\begin{align}
{\cal R}_\text{quad} & = \frac{i}{4\sqrt{2}} D^{\frac{1}{4}} \{ [\bnabla ( D^{-\frac{1}{4}} \chi - D^{-\frac{1}{4}} \chi^*) ]^2 - (D^{\frac{3}{4}} \chi - D^{\frac{3}{4}} \chi^*)^2 \} \label{eq:Rquad} \\
\nonumber & + \frac{i}{2\sqrt{2}} D^{-\frac{1}{4}} \{ \bnabla \cdot [ (D^{\frac{1}{4}} \chi + D^{\frac{1}{4}} \chi^*) \bnabla(D^{-\frac{1}{4}} \chi - D^{-\frac{1}{4}} \chi^*)   ]  \} \\
\nonumber & + \frac{i}{2\sqrt{2}} D^{\frac{3}{4}} \{  (D^{\frac{1}{4}} \chi + D^{\frac{1}{4}} \chi^*) (D^{\frac{3}{4}} \chi - D^{\frac{3}{4}} \chi^*)     \}  \, . 
\end{align}
The cubic term reads:
\begin{align}
 {\cal R}_\text{cubic} & = - \frac{i}{4} D^{\frac{3}{4}} \{ (D^{\frac{1}{4}} \chi + D^{\frac{1}{4}} \chi^*) D\{ (D^{\frac{1}{4}} \chi + D^{\frac{1}{4}} \chi^*)   (D^{\frac{3}{4}} \chi - D^{\frac{3}{4}} \chi^*) \}  \}  \label{eq:fullRcubic} \\
\nonumber & + \frac{i}{8} D^{\frac{3}{4}} \{ (D^{\frac{1}{4}} \chi + D^{\frac{1}{4}} \chi^*)^2    (D^{\frac{7}{4}} \chi - D^{\frac{7}{4}} \chi^*)   \} \\
\nonumber & + \frac{i}{8} D^{\frac{7}{4}} \{ (D^{\frac{1}{4}} \chi + D^{\frac{1}{4}} \chi^*)^2    (D^{\frac{3}{4}} \chi - D^{\frac{3}{4}} \chi^*)   \} \\
\nonumber & - \frac{i}{4} D^{\frac{1}{4}} \{ (D^{\frac{1}{4}} \chi + D^{\frac{1}{4}} \chi^*)    (D^{\frac{3}{4}} \chi - D^{\frac{3}{4}} \chi^*) (D^{\frac{7}{4}} \chi - D^{\frac{7}{4}} \chi^*)   \} \\
\nonumber  & + \frac{i}{4} D^{\frac{1}{4}} \{ (D^{\frac{3}{4}} \chi - D^{\frac{3}{4}} \chi^*) D\{ (D^{\frac{1}{4}} \chi + D^{\frac{1}{4}} \chi^*)    (D^{\frac{3}{4}} \chi - D^{\frac{3}{4}} \chi^*)  \} \} \, . 
\end{align}

While this term has a complicated form, its contribution to the solvability condition takes a much more compact form, as we now discuss.

\subsection{Collecting the resonant terms\label{sec:collecting}}

In this appendix we propose alternative forms for the various terms entering the solvability conditions~(\ref{eq:SCth}). The proposed forms are more compact and they lead to the same solvability conditions. We use the symbol $\hookrightarrow$ to indicate `has the same contribution to the solvability condition as'. 

The resonant terms are proportional to $e^{-it}$ and they enter the solvability condition only through their projection onto $e^{i {\bf k}\cdot {\bf x}}$ for the wavevectors ${\bf k}$ satisfying $k=1+{\cal O}(\delta^2)$. The resonant contribution from the cubic term ${\cal R}_\text{cubic}|_{\chi_1}$ is deduced from expression~(\ref{eq:fullRcubic}). When evaluating this contribution, $\chi$ is replaced by $\chi_1$, $D^{\alpha}{\chi_1}$ is replaced by $\chi_1$, and any $D^{\alpha}$ operator arising in front of a term can be replaced by one, because the solvability condition only extracts signal at unit wavenumber. Isolating the contribution oscillating as $e^{-it}$ and making these simplifications leads to the following equivalent form for the resonant terms arising from ${\cal R}_\text{cubic}|_{\chi_1}$:
\begin{align}
{\cal R}_\text{cubic}|_{\chi_1} & \hookrightarrow \frac{i}{2} \chi_1^* (\chi_1^2- D\{\chi_1^2 \})   \, . \label{eq:cubiccontribution}
\end{align}
The resonant terms in the cross term ${\cal R}_\text{quad}|_{{\chi_1} \leftrightarrow \chi_2}$ arise from the various contributions that oscillate as $e^{-it}$: the cross terms between $\chi_1^*$ and $\chi_2^{(-2)}e^{-2it}$, the cross terms between $\chi_1^*$ and $\chi_2^{(2)*}e^{-2it}$, and the cross terms between $\chi_1$ and $\chi_2^{(0)}$ (together with $\chi_2^{(0)*}=\chi_2^{(0)}$). We denote these contributions as ${\cal R}_\text{quad}|_{ \chi_1^* \leftrightarrow   {\chi_2^{(-2)}} } $, ${\cal R}_\text{quad}|_{  \chi_1^* \leftrightarrow {\chi_2^{(2)*}} }$ and ${\cal R}_\text{quad}|_{ \chi_1 \leftrightarrow {\chi_2^{(0)}}  } $, respectively.
Once again, when evaluating these contributions we can replace $D^\alpha$ by one when it applies to $\chi_1$ or $\chi_1^*$, and we can also replace $D^\alpha$ by one if the operator arises at the very beginning of a term. With these simplifications the contributions from the cross terms to the solvability condition can be recast as:
\begin{align}
{\cal R}_\text{quad}|_{ \chi_1^* \leftrightarrow   {\chi_2^{(-2)}} } & \hookrightarrow \frac{i}{2\sqrt{2}} \left[-\bnabla {\chi_1}^* \cdot \bnabla (D^{\frac{1}{4}} \chi_2^{(-2)} ) + {\chi_1}^* (2D^{\frac{3}{4}}- D^{\frac{7}{4}}) \{ \chi_2^{(-2)} \}  \right] e^{-2it} \, , \\
{\cal R}_\text{quad}|_{  \chi_1^* \leftrightarrow {\chi_2^{(2)*}} } & \hookrightarrow \frac{i}{2\sqrt{2}} \left[-\bnabla {\chi_1}^* \cdot \bnabla (D^{\frac{1}{4}} \chi_2^{(2)*} ) + {\chi_1}^* (D^{\frac{7}{4}}-2D^{\frac{3}{4}}) \{ \chi_2^{(2)*} \}  \right] e^{-2it} \, , \\
{\cal R}_\text{quad}|_{ \chi_1 \leftrightarrow {\chi_2^{(0)}}  } & \hookrightarrow \frac{i}{\sqrt{2}} \bnabla {\chi_1} \cdot \bnabla ( D^{\frac{1}{4}}   \chi_2^{(0)}  )  \, .
\end{align}
Substituting expressions~(\ref{eq:chi2minus2}-\ref{eq:chi2plus2}) for $\chi_2^{(-2)}$, $\chi_2^{(0)}$ and $\chi_2^{(2)}$ and summing the three previous contributions leads to the following simpler form for the contribution from the cross term to the solvability condition:
\begin{align}
{\cal R}_\text{quad}|_{{\chi_1} \leftrightarrow \chi_2} & \hookrightarrow - \frac{i}{8} \bnabla {\chi_1}^* \cdot \bnabla \left[ \frac{D}{D-4} \left\{ (\bnabla {\chi_1})^2 + (3-2D) \{ {\chi_1}^2 \} \right\}  \right]   \label{eq:crosstermtemp} \\
\nonumber & + \frac{i}{8}  {\chi_1}^* \frac{D(2-D)}{D-4} \left\{ 2 (\bnabla {\chi_1})^2 + (-D^2+2D-2) \{ {\chi_1}^2\}  \right\}  \\
\nonumber  & +  \frac{i}{4} \left[ \bnabla {\chi_1} \cdot \bnabla \left( |{\chi_1}|^2 - |\bnabla {\chi_1}|^2   \right)   \right]  \, . 
\end{align}
To further simplify this expression, we use the narrow-band relation $\Delta \chi_1= -\chi_1 + {\cal O}(\delta^3)$ to obtain:
\begin{align}
(\bnabla \chi_1)^2 & =\frac{1}{2} \Delta (\chi_1^2) - \chi_1 \Delta \chi_1 = \frac{1}{2} \Delta (\chi_1^2) + \chi_1^2 + {\cal O}(\delta^4) \\
|\bnabla \chi_1|^2 & =\frac{1}{2} \Delta (|\chi_1|^2) - \frac{1}{2}(\chi_1 \Delta \chi_1^* + \chi_1^* \Delta \chi_1 )  = \frac{1}{2} \Delta (|\chi_1|^2) + |\chi_1|^2 + {\cal O}(\delta^4) \, .
\end{align}
Because we only collect the ${\cal O}(\delta^3)$ terms in the solvability condition, we can substitute these approximate relations into~(\ref{eq:crosstermtemp}), which leads to:
\begin{align}
{\cal R}_\text{quad}|_{{\chi_1} \leftrightarrow \chi_2}  \hookrightarrow & \frac{i}{16} \bnabla {\chi_1}^* \cdot \bnabla \left[ \frac{D(D^2+4D-8)}{D-4}  \{ {\chi_1}^2 \}  \right]  \label{eq:tempquadcontribution} \\
\nonumber & + \frac{i}{4}  {\chi_1}^* \frac{D^2(D-2)(D-1)}{D-4}  \{ {\chi_1}^2\}   -  \frac{i}{8}  \bnabla {\chi_1} \cdot \bnabla \Delta \{ |{\chi_1}|^2  \}      \, . 
\end{align}

We now want to show that the contribution to the solvability condition from the term ${\cal T}_1=-  \frac{i}{8}  \bnabla {\chi_1} \cdot \bnabla \Delta \{ |{\chi_1}|^2  \}$ is the same as that from the combination ${\cal T}_2+{\cal T}_3$, where ${\cal T}_2=i \chi_1^* g_2(D) \{ \chi_1^2 \}$ and ${\cal T}_3=i \chi g_3(D) \{ |\chi|^2 \}$, with $g_2(D)$ and $g_3(D)$ to be determined. Inserting expression~(\ref{eq:solchi1}) for $\chi_1$ into ${\cal T}_1$, we obtain:
\begin{align}
{\cal T}_1= - \frac{i}{8} \sum_{{\bf k}_1,{\bf k}_2,{\bf k}_3} \, A_{{\bf k}_1} A_{{\bf k}_2} A_{{\bf k}_3}^*  {\bf k}_1 \cdot  ({\bf k}_2-{\bf k}_3)  | {\bf k}_2-{\bf k}_3 |^2 e^{i[({\bf k}_1+{\bf k}_2-{\bf k}_3)\cdot {\bf x} -t]} \, , \label{eq:sumT1}
\end{align}
where the sum is over wavevectors of near-unit norm and we omit the slow-time dependence of the coefficients $A_{\bf k_j}(T)$ to alleviate notations. The contribution from ${\cal T}_1$ to the solvability condition~(\ref{eq:SCth}) written for wavevector ${\bf k}$ is nonzero only when ${\bf k}_1+{\bf k}_2-{\bf k}_3={\bf k}$. The wavevectors being all of near-unit norm, they either form a rhombus with ${\bf k}_2 \simeq {\bf k}$ and ${\bf k}_1 \simeq {\bf k}_3$, or ${\bf k}_1 \simeq {\bf k}$ and ${\bf k}_2 \simeq {\bf k}_3$, or they form a wedge, with ${\bf k}_2 \simeq -{\bf k}_1$ and ${\bf k}_3 \simeq -{\bf k}$, see figure~\ref{fig:quartet}. We define the ensembles of horizontal wavevectors ${\cal D}_r$ and ${\cal D}_w$ associated with such resonant quartets, the former with rhombus geometry and the latter with wedge geometry:
\begin{align}
\nonumber {\cal D}_r  = & \{({\bf k}_1,{\bf k}_2,{\bf k}_3) \in \left( \frac{2 \pi}{L} \mathbb{Z}^2 \right)^3, \text{ s.t. } |{\bf k}_1|=1+{\cal O}(\delta^2), |{\bf k}_2|=1+{\cal O}(\delta^2), |{\bf k}_3|=1+{\cal O}(\delta^2),  \\
&   {\bf k}_1={\bf k}_3 + {\cal O}(\delta^2), {\bf k}_2={\bf k} + {\cal O}(\delta^2)  \} \, , \\
\nonumber {\cal D}_w  = & \{({\bf k}_1,{\bf k}_2,{\bf k}_3) \in \left( \frac{2 \pi}{L} \mathbb{Z}^2 \right)^3, \text{ s.t. }|{\bf k}_1|=1+{\cal O}(\delta^2), |{\bf k}_2|=1+{\cal O}(\delta^2), |{\bf k}_3|=1+{\cal O}(\delta^2),  \\
&   {\bf k}_2=-{\bf k}_1 + {\cal O}(\delta^2), {\bf k}_3=-{\bf k} + {\cal O}(\delta^2)  \} \, .
\end{align}
To isolate the resonant part of ${\cal T}_1$,  the sum in equation~(\ref{eq:sumT1}) can be split into $\sum_{{\bf k}_1,{\bf k}_2,{\bf k}_3} =\sum_{({\bf k}_1,{\bf k}_2,{\bf k}_3)\in {\cal D}_r}+\sum_{({\bf k}_2,{\bf k}_1,{\bf k}_3)\in {\cal D}_r} +\sum_{({\bf k}_1,{\bf k}_2,{\bf k}_3)\in {\cal D}_w} $, where the second term corresponds to a permutation of ${\bf k}_1$ and ${\bf k}_2$. For ${\cal T}_1$ the second sum vanishes and we are left with:
\begin{align}
{\cal T}_1 \hookrightarrow \, & i \sum_{({\bf k}_1,{\bf k}_2,{\bf k}_3)\in {\cal D}_r} \, A_{{\bf k}_1} A_{{\bf k}_2} A_{{\bf k}_3}^*  (\cos^4 \theta - 2 \cos^2 \theta +1) e^{i[({\bf k}_1+{\bf k}_2-{\bf k}_3)\cdot {\bf x} -t]} \label{eq:resT1}\\
\nonumber & + i \sum_{({\bf k}_1,{\bf k}_2,{\bf k}_3)\in {\cal D}_w} \, A_{{\bf k}_1} A_{{\bf k}_2} A_{{\bf k}_3}^*  (\cos^4 \theta - 2 \cos^2 \theta +1) e^{i[({\bf k}_1+{\bf k}_2-{\bf k}_3)\cdot {\bf x} -t]} \, .
\end{align}
To obtain the expression above, we have substituted the near equalities between the different wavevectors in the ensemble ${\cal D}_r$ or ${\cal D}_w$, neglecting the ${\cal O}(\delta^2)$ corrections, and we have introduced the half angle $\theta$ between ${\bf k}_1$ and ${\bf k}$, such that ${\bf k}=(\cos \theta, \sin \theta)$ and ${\bf k}_1=(\cos \theta, -\sin \theta)$, see figure~\ref{fig:quartet}.

We now consider the term ${\cal T}_2$, in which we insert expression~(\ref{eq:solchi1}) for $\chi_1$:
\begin{align}
{\cal T}_2= i \sum_{{\bf k}_1,{\bf k}_2,{\bf k}_3} \, A_{{\bf k}_1} A_{{\bf k}_2} A_{{\bf k}_3}^* g_2(|{\bf k}_1+{\bf k}_2|)  e^{i[({\bf k}_1+{\bf k}_2-{\bf k}_3)\cdot {\bf x} -t]}  \, .
\end{align}
Splitting the sum into the three resonant sets leads to:
\begin{align}
{\cal T}_2 \hookrightarrow \, & i \sum_{({\bf k}_1,{\bf k}_2,{\bf k}_3)\in {\cal D}_r} \, A_{{\bf k}_1} A_{{\bf k}_2} A_{{\bf k}_3}^*  g_2(|{\bf k}+{\bf k}_1|) e^{i[({\bf k}_1+{\bf k}_2-{\bf k}_3)\cdot {\bf x} -t]} \\
& i \sum_{({\bf k}_2,{\bf k}_1,{\bf k}_3)\in {\cal D}_r} \, A_{{\bf k}_1} A_{{\bf k}_2} A_{{\bf k}_3}^*  g_2(|{\bf k}+{\bf k}_2|) e^{i[({\bf k}_1+{\bf k}_2-{\bf k}_3)\cdot {\bf x} -t]} \\
& + i \sum_{({\bf k}_1,{\bf k}_2,{\bf k}_3)\in {\cal D}_w} \, A_{{\bf k}_1} A_{{\bf k}_2} A_{{\bf k}_3}^* g_2(0) e^{i[({\bf k}_1+{\bf k}_2-{\bf k}_3)\cdot {\bf x} -t]} \, .
\end{align}
We swap the dummy variables ${\bf k}_1$ and ${\bf k}_2$ in the second sum, which leads to a contribution equal to the first sum:
\begin{align}
{\cal T}_2 \hookrightarrow \, & i \sum_{({\bf k}_1,{\bf k}_2,{\bf k}_3)\in {\cal D}_r} \, A_{{\bf k}_1} A_{{\bf k}_2} A_{{\bf k}_3}^*  2 g_2(2|\cos \theta|) e^{i[({\bf k}_1+{\bf k}_2-{\bf k}_3)\cdot {\bf x} -t]} \label{eq:resT2} \\
\nonumber & + i \sum_{({\bf k}_1,{\bf k}_2,{\bf k}_3)\in {\cal D}_w} \, A_{{\bf k}_1} A_{{\bf k}_2} A_{{\bf k}_3}^* g_2(0) e^{i[({\bf k}_1+{\bf k}_2-{\bf k}_3)\cdot {\bf x} -t]} \, .
\end{align}
Finally, we proceed in a similar way for ${\cal T}_3$. After inserting expression~(\ref{eq:solchi1}) for $\chi_1$ into ${\cal T}_3$, splitting the sum into the three resonant sets, and swapping dummy wavevectors ${\bf k}_1$ and ${\bf k}_2$ in the second sum, we obtain:
\begin{align}
{\cal T}_3 \hookrightarrow \, & i \sum_{({\bf k}_1,{\bf k}_2,{\bf k}_3)\in {\cal D}_r} \, A_{{\bf k}_1} A_{{\bf k}_2} A_{{\bf k}_3}^*  [g_3(0) +g_3(2|\sin \theta|)] e^{i[({\bf k}_1+{\bf k}_2-{\bf k}_3)\cdot {\bf x} -t]} \label{eq:resT3} \\
\nonumber & + i \sum_{({\bf k}_1,{\bf k}_2,{\bf k}_3)\in {\cal D}_w} \, A_{{\bf k}_1} A_{{\bf k}_2} A_{{\bf k}_3}^* g_3(2 |\cos \theta|) e^{i[({\bf k}_1+{\bf k}_2-{\bf k}_3)\cdot {\bf x} -t]} \, . 
\end{align}
Adding the contributions~(\ref{eq:resT2}) and~(\ref{eq:resT3}) from ${\cal T}_2$ and ${\cal T}_3$ leads to:
\begin{align}
\nonumber {\cal T}_2 + {\cal T}_3 \hookrightarrow \, & i \sum_{({\bf k}_1,{\bf k}_2,{\bf k}_3)\in {\cal D}_r} \, A_{{\bf k}_1} A_{{\bf k}_2} A_{{\bf k}_3}^*  [g_3(0) +g_3(2|\sin \theta|)+2 g_2(2|\cos \theta|)] e^{i[({\bf k}_1+{\bf k}_2-{\bf k}_3)\cdot {\bf x} -t]}  \\
 & + i \sum_{({\bf k}_1,{\bf k}_2,{\bf k}_3)\in {\cal D}_w} \, A_{{\bf k}_1} A_{{\bf k}_2} A_{{\bf k}_3}^* [g_3(2 |\cos \theta|)+g_2(0) ] e^{i[({\bf k}_1+{\bf k}_2-{\bf k}_3)\cdot {\bf x} -t]} \, .  \label{eq:resT2T3}
\end{align}
The rhs corresponds to the same contribution as that from ${\cal T}_1$ (that is, it is equal to the rhs of equation(\ref{eq:resT1})), if we choose the functions $g_2$ and $g_3$ such that:
\begin{align}
g_3(0) +g_3(2|\sin \theta|)+2 g_2(2|\cos \theta|) & = \cos^4 \theta - 2 \cos^2 \theta +1 \, , \label{eq:constraint1}\\
g_3(2 |\cos \theta|)+g_2(0)  & =  \cos^4 \theta - 2 \cos^2 \theta +1 \, . \label{eq:constraint2}
\end{align}
Adding a constant ${\cal C}$ to $g_2$ and subtracting the same constant ${\cal C}$ to $g_3$ does not affect the combination $\nonumber {\cal T}_2 + {\cal T}_3$, and therefore we can impose $g_2(0)=0$. Equation~(\ref{eq:constraint2}) then corresponds to:
\begin{align}
g_3(D)=\frac{D^4}{16}- \frac{D^2}{2}+1 = \left(\frac{D^2}{4} -1\right)^2\, ,
\end{align}
and after substitution into~(\ref{eq:constraint1}),
\begin{align}
g_2(D)=- \frac{D^2}{4}\, .
\end{align}
With these expressions for $g_2(D)$ and $g_3(D)$ at hand, we substitute ${\cal T}_2+{\cal T}_3$ for ${\cal T}_1$ in equation~(\ref{eq:tempquadcontribution}). After re-arranging we obtain:
\begin{align}
{\cal R}_\text{quad}|_{{\chi_1} \leftrightarrow \chi_2}  \hookrightarrow & \frac{i}{16} \bnabla {\chi_1}^* \cdot \bnabla \left[ \frac{D(D^2+4D-8)}{D-4}  \{ {\chi_1}^2 \}  \right]  \label{eq:overallquadcontribution} \\
\nonumber & + \frac{i}{4}  {\chi_1}^*  \frac{D^2(D^2-4D+6)}{D-4}  \{ {\chi_1}^2\}   +i \chi_1  \left(\frac{D^2}{4} -1\right)^2 \{ |\chi_1|^2  \}  \, . 
\end{align}
The operator ${\cal N}\{ \chi_1 \}$ in the main text corresponds to the negative of the sum of~(\ref{eq:cubiccontribution}) and~(\ref{eq:overallquadcontribution}).

\bibliographystyle{jfm}

\bibliography{2way}

\end{document}